\documentclass[twocolumn]{aastex63}

\definecolor{deepgreen}{RGB}{20, 130, 22}

\usepackage{savesym}
\usepackage{newtxtext,newtxmath}
\usepackage{amsmath, amssymb, graphics, bm, graphicx}
\allowdisplaybreaks

\newcommand{\mathsym}[1]{{}}
\newcommand{\unicode}[1]{{}}

\newcommand{\ag}{\alpha}

\newcommand{\cg}{\gamma}

\newcommand{\dg}{\delta}

\newcommand{\Om}{\Omega}
\newcommand{\om}{\omega}

\newcommand{\der}{{\rm d}}

\newcommand{\vphi}{\varphi}

\newcommand{\hx}{{\bm {\hat x}}}
\newcommand{\hy}{{\bm {\hat y}}}
\newcommand{\hz}{{\bm {\hat z}}}

\renewcommand{\hom}{{\bm {\hat \omega}}}

\newcommand{\hn}{{\bm {\hat n}}}

\newcommand{\hp}{{\bm {\hat p}}}
\newcommand{\hmf}{{\bm {\hat m}}_f}
\newcommand{\bom}{{\bm \omega}}

\newcommand{\bcdot}{{\bm \cdot}}
\newcommand{\btimes}{{\bm \times}}

\newcommand{\G}{{\rm G}}
\newcommand{\s}{{\rm s}}

\newcommand{\be}{\begin{equation}}
\newcommand{\ee}{\end{equation}}

\newcommand{\frb}{FRB~180916.J0158+65}
\newcommand{\mr}{\mathrm}

\submitjournal{AJ}

\shorttitle{Periodic FRB Emission from Magnetar Motion}
\shortauthors{Li \& Zanazzi}
\graphicspath{{./}{figures/}}

\begin{document}

\title{Emission Properties of Periodic Fast Radio Bursts from the Motion of Magnetars: Testing Dynamical Models}

\correspondingauthor{Dongzi Li}
\email{dzli@cita.utoronto.ca}

\author{Dongzi Li}
\affiliation{Canadian Institute for Theoretical Astrophysics,
University of Toronto,
60 St. George Street,
Toronto, Ontario, M5S 1A7, Canada}
\affiliation{Department of Physics, University of Toronto, 60 St. George Street, Toronto, ON M5S 1A7, Canada}
\affiliation{Cahill Center for Astronomy and Astrophysics, California Institute of Technology, 
 1216 E California Blvd, Pasadena, CA 91125, US}
\author{J.~J. Zanazzi}
\affiliation{Canadian Institute for Theoretical Astrophysics,
University of Toronto,
60 St. George Street,
Toronto, Ontario, M5S 1A7, Canada}

\begin{abstract}
Recent observations of the periodic Fast Radio Burst source 180916.J0158+65 (FRB 180916) find small linear polarization position angle swings during and between bursts, with a burst activity window that becomes both narrower and earlier at higher frequencies.  Although the observed chromatic activity window disfavors models of periodicity in FRB 180916 driven by the occultation of a neutron star by the optically-thick wind from a stellar companion, the connection to theories where periodicity arises from the motion of a bursting magnetar remains unclear.  In this paper, we show how altitude-dependent radio emission from magnetar curvature radiation, with bursts emitted from regions which are asymmetric with respect to the magnetic dipole axis,  can lead to burst activity windows and polarization consistent with the recent observations.  In particular, the fact that bursts arrive systematically earlier at higher frequencies disfavors theories where the FRB periodicity arises from forced precession of a magnetar by a companion or fallback disk, but is consistent with theories where periodicity originates from a slowly-rotating or freely-precessing magnetar.  Several observational tests are proposed to verify/differentiate between the remaining theories, and pin-down which theory explains the periodicity in FRB 180916.

\end{abstract}

\keywords{
radiation mechanisms: general -- polarization -- stars: neutron -- stars: magnetars
}



\section{Introduction}
Fast radio bursts (FRBs) are short ($\sim\mu$s to $\sim$ms duration) radio bursts with unknown origins. A decade after the first discovery \citep{lorimer2007}, 
massive progress has been made in understanding the nature of FRBs. The discovery repeating FRBs \citep[e.g.][]{2014Spitler,2016Spitler,CHIME2019c,Fonseca+2020} suggests at least a portion of the FRB population has a non-catastrophic origin. The first localization of an FRB source \citep{chatterjee2017} confirmed their cosmological origin. The detection of extremely bright radio bursts from a galactic magnetar \citep{20SGRCHIME,20SGRSTARE2} places magnetars as a most promising progenitor for FRBs.

Repeating FRBs offer great opportunities for follow-up observations, as well as studying the burst properties with time and frequency. 
\frb\ (hereafter FRB 180916) is the closest localized extragalactic FRB source ($z = 0.0337$; \citealt{mnh+20}), and also the most active repeating source detected within the frequency band 400-800 MHz 
by the Canadian Hydrogen Intensity Mapping Experiment
\citep[CHIME,][]{chime2019b}. 
Remarkably, a periodicity in the activity of FRB 180916  has recently been found. Bursts detected at 400-800 MHz occur within a 5-day window that repeats every 16.3 days \citep{chime2020a}. 
Multiple models have been proposed to explain the periodicity: the orbital motion of the FRB source (neutron star or magnetar) around a windy companion (\citealp{16DaiOrbit}; \citealp*{lyutikov+20,20IokaOrbit}), 
 precession of the burst emitting object itself \citep[e.g.][]{20Zanazzi,levin+2020,20YangPrecession}, or ultralong rotational periods of the bursting object \citep{Beniamini2020a}.  
 
Multi-wavelength follow-ups of this source provide constraints on the FRB progenitor and the local environment, and also introduce additional puzzles. 
Imaging from the \textit{Hubble Space Telescope} shows that the location of FRB 180916 is offset by $\sim250$ pc from the nearest knot of star formation in the host galaxy \citep{2020Tendulkar} , while young magnetars are expected to born in a star forming region. 
For bursts detected in the $L$-band, the change of linear polarization position angle (PA) is constrained to be $\lesssim 10^\circ$-$20^\circ$ across the bursts of the same phase \citep{Nimmo2020}, and $\lesssim 50^\circ$ across the $L$-band active phases \citep{20ApertifR3}. No PA swing is observed, as would generically be expected in the magnetospheric origin models (as oppose to the diverse polarization angle swing detected in FRB 180301, \citealt{Luo+2020}). 
Most strikingly, the active phase is observed to be chromatic, with the activity window being both narrower and earlier at higher frequencies \citep{20ApertifR3,20LofarR3}. 
This disfavours models explaining the periodic activity with the eclipse of a companion wind, as these theories predict a narrower activity window at lower frequencies \citep{lyutikov+20,20IokaOrbit}. 

Recently, a period of $\sim$160 days was detected in the repeating FRB 121102 \citep{Rajwade+2020,Cruces+2021}.  Similar to FRB 180916, FRB 121102 also has small PA swings during and between bursts \citep{Michilli+2018}, but data on the frequency-dependence of the activity window has yet to be gathered.  Clearly, the polarization emission and frequency-dependence expected from different models of periodic FRBs is becoming highly topical.

The goal of this work is to show how the recent constraints on the PA variation, as well as the dependence on the FRB activity window with frequency, fit into theories which argue the periodicity of FRB 180916 originates from the motion of a Neutron Star (NS) or magnetar.  Section~\ref{sec:emit} justifies our phenomenological emission model for NSs emitting FRBs.  Section~\ref{sec:dyn} discusses the three dynamical theories which have been put forth to explain the periodicity of FRB 180916, which we test in this work: a NS with a long rotation period, a NS undergoing free precession, and a NS undergoing forced precession.  Section~\ref{sec:result} presents the main results of this work, and discusses which dynamical models are favored to explain the periodicity of FRB 180916.  Section~\ref{sec:obs} discusses future observations which can potentially distinguish between the remaining theories.  Section~\ref{sec:conc} summarizes our work.


\section{Emission Model for FRBs}
\label{sec:emit}

\begin{figure}
    \centering
    \includegraphics[width=\linewidth]{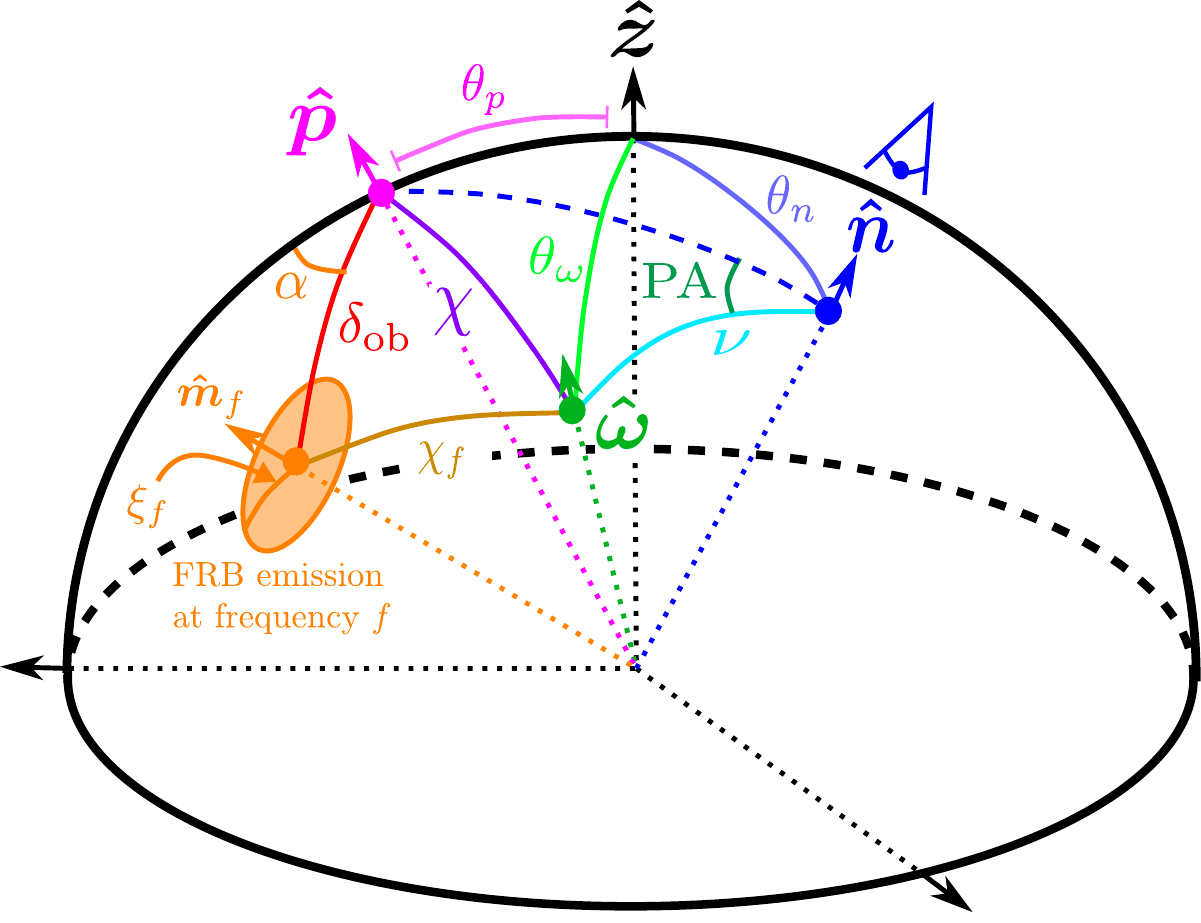}
    \caption{Geometry of our model for FRB emission in the upper-half of a NS's magnetosphere.  The FRB emission at frequency $f$ originates from a cone centered around $\hmf$, with opening angle $\xi_f$.  See \S\ref{sec:emit} for details.}
      \label{fig:setup}
\end{figure}

To make testable predictions for periodic FRB emission, we construct a phenominological model of radio emission from a rotating (and possibly precessing) NS, motivated by models for radio pulsar emission \citep[e.g.][]{RudermanSutherland75,Rankin93}.  Figure~\ref{fig:setup} displays the geometry of the model.  A NS with rotation axis $\hom$ and dipole moment unit vector $\hp$ is viewed by an observer, with $\hn$ being the unit vector pointing in the direction of the Line of Sight (LOS) of the observer.  FRBs are emitted in a direction $\hmf$ with frequency $f$, and an observer sees the FRB emission when $\hn$ lies within the cone centered on $\hmf$ with opening angle $\xi_f$ (orange region of Fig.~\ref{fig:setup}).  The emission axis $\hmf$ is offset from $\hp$ by a magnetic polar angle $\dg_{\rm ob}$ (angle between $\hmf$ and $\hp$), and a magnetic longitude $\ag$ (angle between planes spanned by vector pairs $\{\hx,\hp\}$ and $\{\hmf,\hp\}$).  The magnetic (angle between $\hp$ and $\hom$) and emission (angle between $\hmf$ and $\hom$) angles are $\chi$ and $\chi_f$, respectively, while the angle between $\hn$ and $\hom$ is $\nu$.  The angles between $\hp$, $\hom$, and $\hn$, and the symmetry axis $\hz$, are $\theta_p$, $\theta_\om$, and $\theta_n$, respectively.  The position angle ${\rm PA}$ of the linear polarization (in the rotating vector model, \citealt{RadhakrishnanCooke69}) is the angle between the planes spanned by the vector pairs $\{\hn,\hom\}$ and $\{\hn,\hp\}$.

The emission model assumed in this paper is significantly more complex than the model in \cite{20Zanazzi}, which assumed $\hmf = \hp$.  In this section, we justify our model assumptions about the FRB emission region, assuming the coherent radio emission arises due to curvature radiation from charged particles travelling along the NS magnetic field lines.  Magnetar curvature radiation has already been invoked by many models to explain the coherent radio emission characteristic of FRBs \citep[e.g.][]{Katz2014,CordesWasserman2016,Kumar+2017,LuKumar2018,YangZhang18,Lu+2020,Yang+2020} Section~\ref{sec:emit_dir} discusses our expectations for how the emission direction $\hmf$ changes with FRB frequency $f$, while Section~\ref{sec:emit_pol} discusses the linear polarization of the emission.

\subsection{FRB Emission Direction}
\label{sec:emit_dir}

  \begin{figure}
    \centering
    \includegraphics[width=\linewidth]{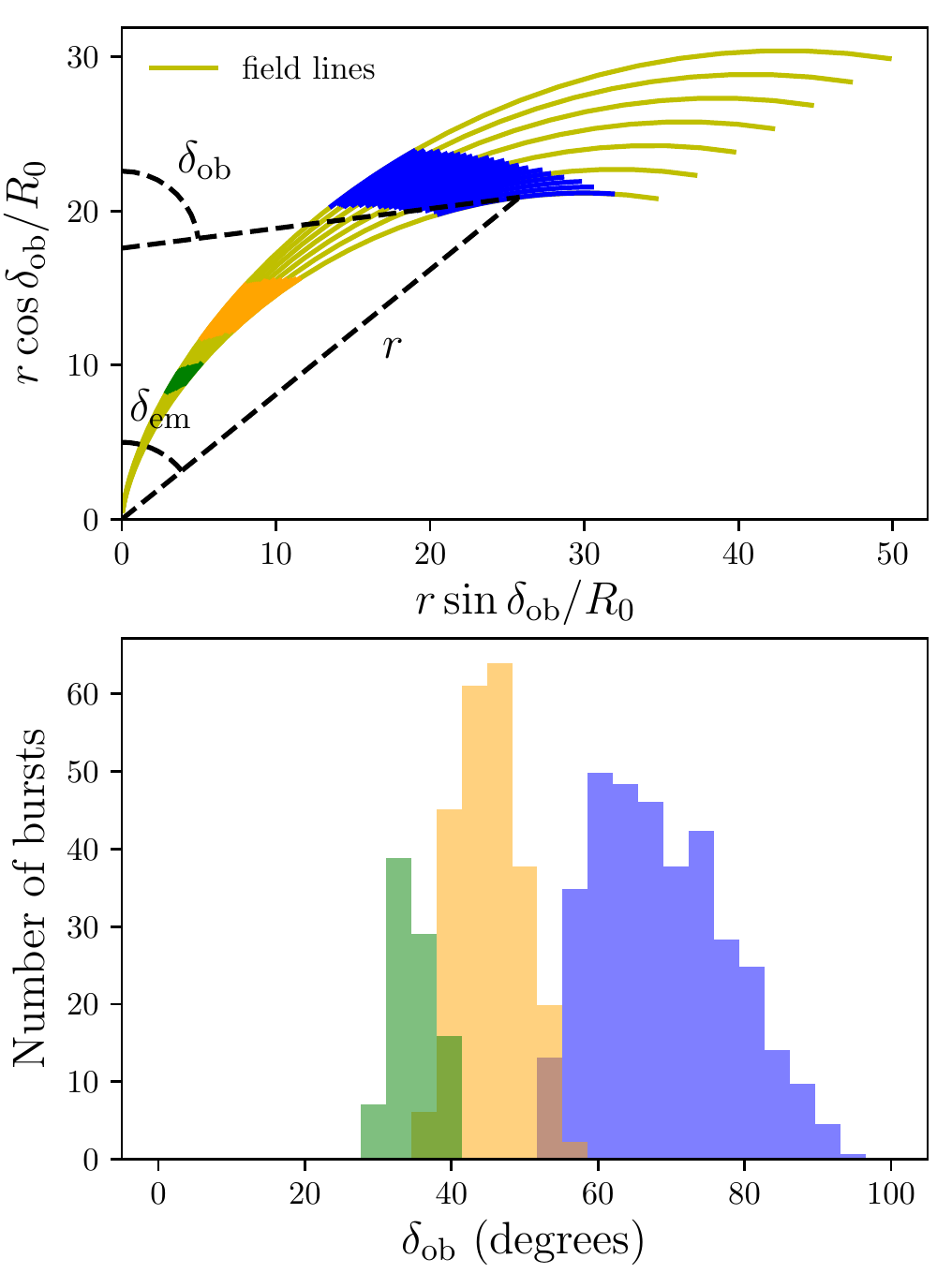}
    \caption{
    Altitude and angular dependence of FRB emission from curvature radiation (eq.~[\ref{eq:f2r}]), for $55 \le C \le 80$.  \textit{Top panel:} Trajectories of magnetic field lines (eq.~[\ref{eq:dipole}]).  Field lines which cause emission in different frequency bands are colored green (1.36-2 GHz), orange (400-800 MHz), and blue (110-180 MHz).  The magnetic polar angles $\dg_{\rm em}$ and $\dg_{\rm ob}$ are defined by the black dashed lines.  \textit{Bottom panel:} Histogram of observed magnetic polar angle $\dg_{\rm ob}$ of bursts from a centered dipole magnetic field.  The histogram is computed assuming an equal number of bursts are emitted per unit length of the field line, with $C$ spaced linearly, for the frequency bands 1.36-2 GHz (green), 400-800 MHz (orange), and 110-180 MHz (blue).  The normalization for the Number of Bursts is arbitrary.  The altitude of the burst emission, as well as the mean and variance of the burst emission $\dg_{\rm ob}$ values, increase with decreasing burst frequency $f$.
    }
       \label{fig:polarangle}
\end{figure}

\begin{figure}
    \centering
    \includegraphics[width=\linewidth]{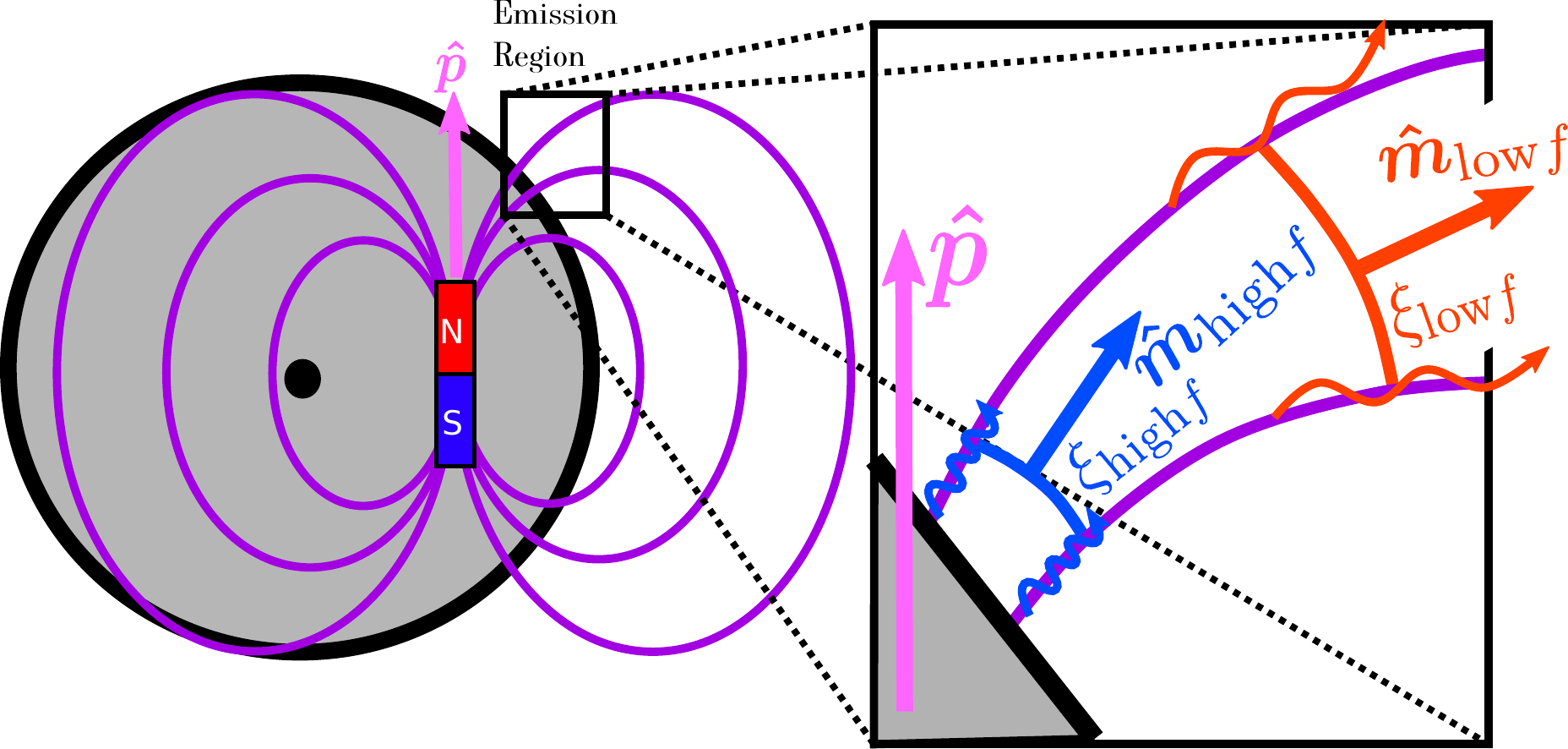}
    \caption{
    Cartoon representation of offset magnetic field geometry, which can give rise to FRB emission from curvature radiation consistent with our model (Fig.~\ref{fig:setup}).  The strong exposed magnetic fields on one side of the NS lead to burst emission occuring in ``cones'' around the emission axis $\hmf$, with both the magnetic polar angle $\dg_{\rm ob}$ (angle between $\hmf$ and $\hp$) and opening angle $\xi_f$ increasing with decreasing frequency $f$.
    }
       \label{fig:offaxis}
\end{figure}

In the study of pulsar emission, it is widely believed that different frequencies are produced at different altitudes within the pulsar magnetosphere (e.g. \citealt{1977puls.book.Manchester}).  It has been shown the spread of field lines at higher altitude could account for the widening of pulse profiles at lower frequencies (e.g. \citealt{1978Cordes.radius2freq,1992Phillips.radius2freq}). 
This kind of ``radius-to-frequency mapping'' has also been used to explain the downward drifting pattern seen in some repeating FRBs \citep{19WangDownwardDrift, Lyutikov2020}.  Here, we demonstrate this same effect in the context of curvature radiation, and show how it 
will lead to the magnetic polar angle $\dg_{\rm ob}$ and cone opening angle $\xi_f$ of the observed FRB emission to decrease with an increase in frequency $f$.

For simplicity, we assume the magnetic field exterior to the NS is a dipole in vacuum, and ignore how plasma affects the magnetic field \citep[e.g.][]{GoldreichJulian1969,Tchekhovskoy+2016,PhilippovSpitkovsky2018}.  The trajectory of a single field line in polar coordinates is given by
\begin{equation}
    \label{eq:dipole}
    \frac{r}{R_0}= C\sin^2\delta_\mr{em}
\end{equation}
where $\dg_{\rm em}$ is the magnetic polar angle of the field line, $r$ is the distance to the center of the NS, $R_0$ is the NS radius, while $1 \le C < \infty$ is a constant which varies for different field lines.  Emission at $\dg_{\rm em}$ will be observed by a distant observer at a different magnetic polar angle $\dg_{\rm ob}$, which is related to $\dg_{\rm em}$ by
\begin{equation}
\label{eq:delta}
    \cos\delta_\mr{ob}=\frac{1+3\cos2\delta_\mr{em}}{\sqrt{10+6\cos2\delta_\mr{em}}}.
\end{equation}
The trajectory of a number of different magnetic field lines, as well as the relation between $\dg_{\rm em}$ and $\dg_{\rm ob}$, are displayed in the top panel of Figure~\ref{fig:polarangle}.

The characteristic frequency $f$ of emission from curvature radiation is \citep[e.g.][]{RudermanSutherland75}:
\begin{equation}
	 f=\frac{3}{4\pi}\gamma^3 \frac{c}{\rho},
	 \label{eq:f_curve}
\end{equation}
where $\cg$ is the Lorentz factor, while $\rho$ is the curvature radius.  For a dipolar field (eq.~[\ref{eq:dipole}]), $\rho$ can be shown to be \citep[e.g.][]{YangZhang18,19WangDownwardDrift}
\begin{equation}
\label{eq:rho}
    \rho = \frac{r(1+3\cos^2{\delta_\mr{em}})^{\frac{3}{2}}}{3\sin{\delta_\mr{em}}(1+\cos^2\delta_\mr{em})}
    \equiv r F(\dg_{\rm em}).
\end{equation}
Assuming for our simple model, the Lorentz factor decreases with $r$ as $\gamma(r)=\gamma_0 (r/R_0)^{-2/3}$ (also assumed in e.g. \citealt{Lyutikov2020} to explain the downward drifting rate of FRBs), the emission frequency becomes
\begin{equation}
\label{eq:f2r}
			f=K_0 F(\dg_{\rm em})\, \left(\frac{r}{R_0}\right)^{-3}
\end{equation}
where $K_0 = 3c\gamma_0^3/(4\pi R_0)$, with $R_0 = 10^6 \, {\rm cm}$ and $K_0 = 5000 \, {\rm GHz}$ for our model (corresponding to $\cg_0 \approx 900$)\footnote{Notice that in Equation~\ref{eq:f2r}, a change from $r$ to $r^\prime$ can be compensated by a change of $K_0$ to $K_0 r^3/r^{\prime3}$ to emit in the same frequency. Therefore, the emission height $r$ in Figure~\ref{fig:polarangle} can be scaled with a change of $\cg_0$.}.  

The upper panel of Figure~\ref{fig:polarangle} shows the emission heights of different frequency bands, displayed in green ($1.36-2 \, {\rm GHz}$), orange ($400-800 \, {\rm MHz}$), and blue ($110-180 \, {\rm MHz}$).  Clearly for curvature radiation, high (low) frequency radiation originates at low (high) altitudes.  The bottom panel displays a histogram for number of bursts with a given $\dg_{\rm ob}$ value, assuming an equal number of bursts are emitted per unit length of the field line, with the field line $C$ values spaced linearly between $55 \le C \le 80$.  From this, we see both the mean and spread of $\dg_{\rm ob}$ values for a given frequency band increases with decreasing frequency.

Within this simple model, a dipole situated at the center of a NS will lead to FRB emission symmetric about the NS dipole moment $\hp$, rather than localized on a small ``patch'' at a location displaced from $\hp$.  One simple modification to the magnetic field geometry which can lead to asymmetric emission about the $\hp$ axis is displayed in Figure~\ref{fig:offaxis}: a dipole displaced from the NS center.  The strong magnetic fields exposed on one side of the NS (which are buried on the other side in this cartoon) can lead to emission ``cones'' similar to those assumed in our model (Fig.~\ref{fig:setup}), with opening angles comparable to the width of the histogram widths displayed in the bottom panel of Figure~\ref{fig:polarangle}.  There is growing observational evidence many NSs may have similar magnetic field geometries as displayed in Figure~\ref{fig:offaxis}.  An offset dipole has been invoked to explain X-ray emission from the mode-switching pulsar PSR B0943+10 \citep{Storch14}.  Recently, X-ray observations from the NICER mission found the hot spots on the surface of isolated pulsars to be far from antipodal, implying a highly complex magnetic field far from the classic assumption of a centered dipole \citep{Riley2019,Bilous2019a}.  For this work, rather than attempting to construct a complex magnetic field which can lead to FRB emission consistent with our model, we simply leave $\dg_{\rm ob}$ and $\xi_f$ as free parameters, with the general expectation both $\dg_{\rm ob}$ and $\xi_f$ should increase with decreasing $f$.

\subsection{FRB Polarization}
\label{sec:emit_pol}

To model the polarization of FRB emission, we use the rotating vector model \citep{RadhakrishnanCooke69}, where the position angle ${\rm PA}$ of the linear polarization is given by
\be
\tan {\rm PA} = \frac{-\sin \Psi \sin \chi}{\cos \chi \sin \nu - \cos \nu \sin \chi \cos \Psi},
\label{eq:RVM}
\ee
where $\Psi$ is the rotational phase ($\Psi = 0$ corresponds to when $\hn$ has its closest approach to $\hp$), while the other quantities are displayed in Figure~\ref{fig:setup}.  Although the magnetic field close to the NS may be complex, \cite{Lu+19} argued the propagation of a FRB across a plasma-filled NS magnetosphere causes the electric field of the burst to ``freeze'' in a direction perpendicular to the magnetic field when the plasma density is sufficiently low (and the magnetic field is approximately dipolar), implying equation~\eqref{eq:RVM} should be an adequate model for the linear polarization from FRBs.  Other works have also used PA measurements constrain the free-precession of NSs \citep{Weisberg+2010}.  We neglect how additional propagation effects can cause the polarization to differ from the rotating vector model \citep[e.g.][]{Wang+10,BeskinPhilippov12}.

\section{Dynamics of Periodic FRB Models}
\label{sec:dyn}

\begin{figure*}
	\begin{minipage}{0.3\linewidth}
    \hspace{-0.1cm}
    \includegraphics[width=\linewidth]{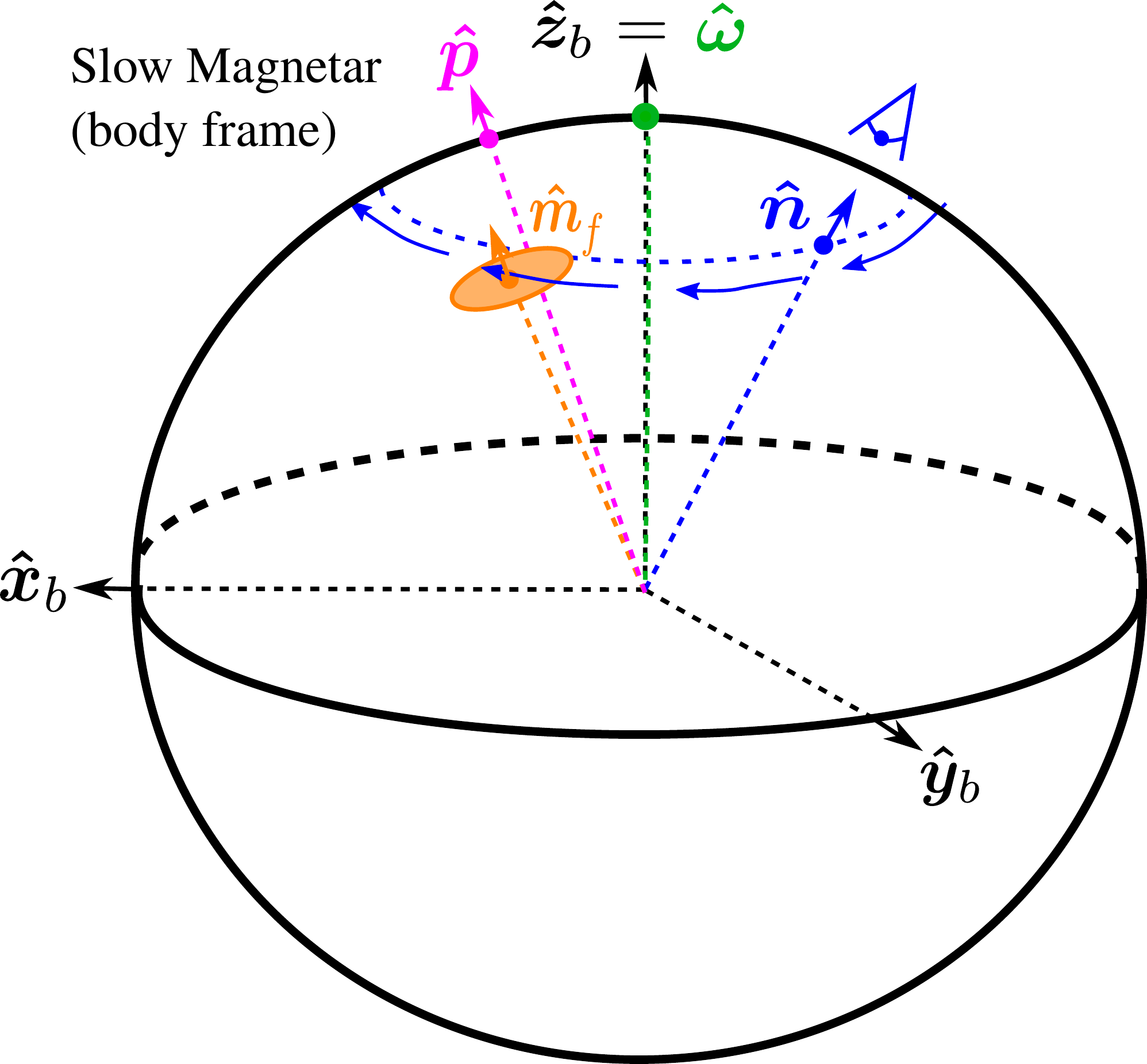}
    \end{minipage}
        \hfill
	\begin{minipage}{0.3\linewidth}
    \includegraphics[width=\linewidth]{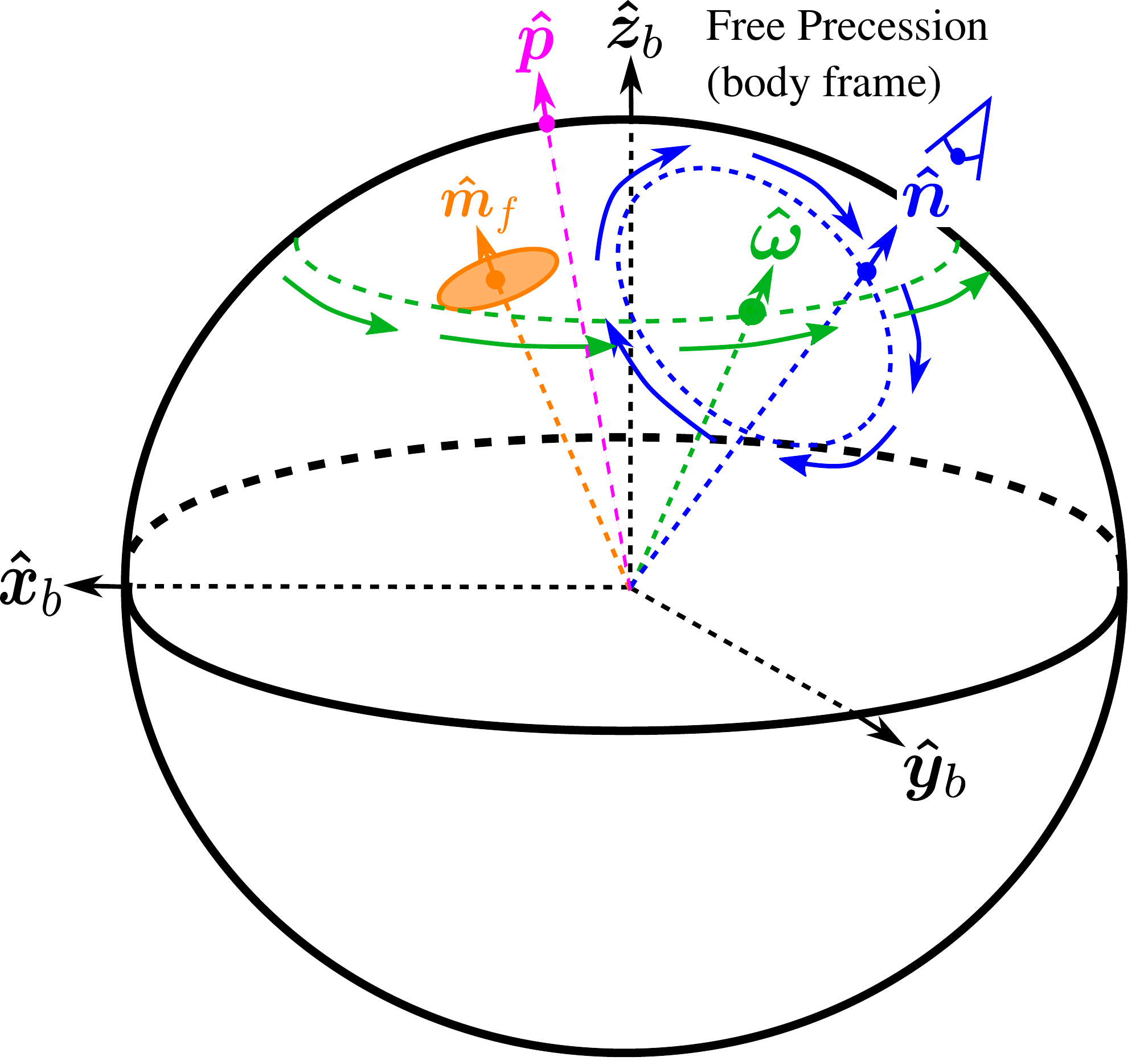}
    \end{minipage}
        \hfill
        	\begin{minipage}{0.3\linewidth}
    \includegraphics[width=\linewidth]{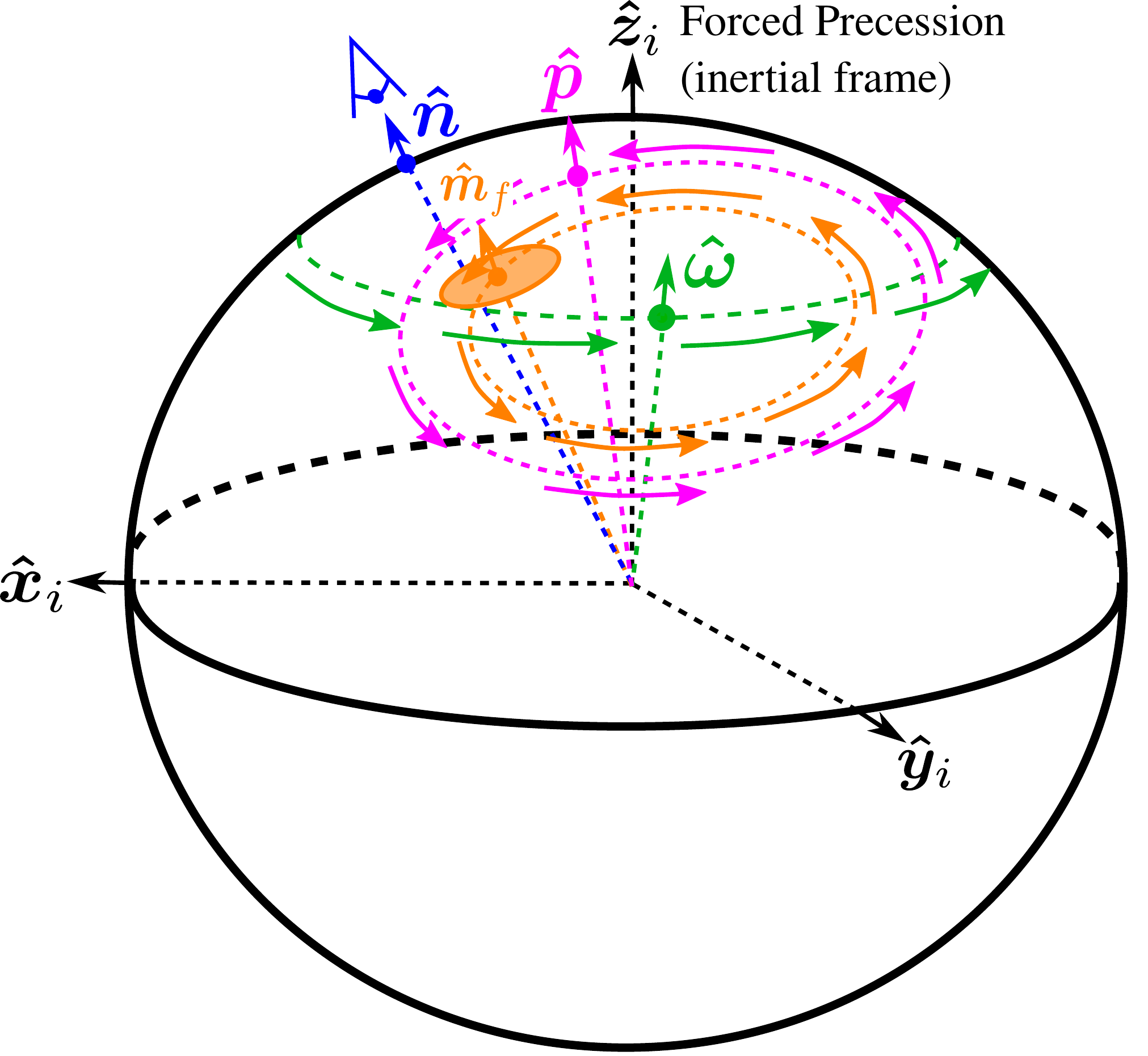}
    \end{minipage}
    \caption{Dynamics of three different periodic FRB models.  Here, $\hom$ is the spin axis, $\hn$ is a unit vector in the the LOS direction, $\hp$ is the dipole moment unit vector, and $\hmf$ represents a unit vector pointing in the direction of the FRB emission. \textit{Left panel:} The slowly rotating magnetar model.  Here $\hom$ and $\hmf$ are fixed in the body frame $\{\hx_b,\hy_b,\hz_b\}$, while $\hn$ rotates around $\hom$. The coordinates are defined by $\hz_b = \hom$, and $\hx_b$ by the projection of $\hp$ onto the plane perpendicular to $\hz_b$.
    \textit{Middle panel:} The free precession model.  Here, the precession axis $\hz_b$,  as well as $\hp$ and $\hmf$, are fixed in the body frame $\{\hx_b, \hy_b, \hz_b\}$, while $\hom$  precesses around $\hz_b$, while $\hn$ rotates around $\hom$.  The coordinates are defined by $\hz_b$ being the axis $\hom$ precesses about in the body frame, while $\hx_b$ defined by the projection of $\hp$ onto the plane perpendicular to $\hz_b$.
    \textit{Right panel:} Forced precession model.  Here, the precession axis $\hz_i$ is fixed in the system's inertial frame $\{\hx_i,\hy_i,\hz_i\}$ (which we work within) instead of body frame $\{\hx_b, \hy_b, \hz_b\}$. In the inertial frame, $\hn$ is fixed, while $\hom$ precesses around the precession axis $\hz_i$, with $\hmf$ and $\hp$ rotating around $\hom$.  The coordinates are defined by $\hz_i$ being the axis $\hom$ precesses about, while $\hx_i$ defined by the projection of $\hn$ onto the plane perpendicular to $\hz_i$.}
   \label{fig:dyn_models}
  \end{figure*}
  
Models which ascribe the periodicity of FRB 180916 from the motion of a magnetar fall into three different categories.  The simplest dynamical model postulated the 16.3 day periodicity was the rotation period of the magnetar \citep{Beniamini2020a}, which implies this magnetar must have a rotation period much longer than those typically observed ($\sim$0.1-10 s, e.g. \citealt{KaspiBeloborodov2017}).  Models which assumed more typical magnetar spin periods argued the periodicity of FRB 180916 came from either free (or ``inertial'') precession of the NS \citep{20Zanazzi,levin+2020,Sobyanin2020}, or forced precession of the NS from a companion \citep{Yang2020} or a fallback disk \citep{Tong+2020}.  In this section, we show how each scenario predicts different motion of a bursting magnetar with respect to a distant observer.  Figure~\ref{fig:dyn_models} summarizes how the motion of the NS spin axis $\hom$, dipole moment axis $\hp$, emisison direction axis $\hmf$, and observer LOS axis $\hn$ differ between the three different dynamical theories.  We defer a discussion of the physics which lead to these three different classes of magnetar motions to the references above.

The dynamics of these different motions are more conveniently analyzed in either a Cartesian coordinate system anchored into and co-rotating with the NS $\{\hx_b,\hy_b,\hz_b\}$ (body frame), or stationary with respect to a distant observer $\{\hx_i,\hy_i,\hz_i\}$ (inertial frame).  The time evolution of a vector ${\bm v}(t)$ in the body $\der {\bm v}/\der t|_b$ or inertial $\der {\bm v}/\der t|_i$ frames are related via $\der {\bm v}/\der t|_b + \bom \btimes {\bm v} = \der {\bm v}/\der t|_i$, where $\bom = \om \hom$, with $\om$ the spin frequency of the NS.

\subsection{Magnetar with Slow Rotation}
\label{sec:dyn_rot}

The left panel of Figure~\ref{fig:dyn_models} displays our dynamical model for periodic FRBs due to slowly-rotating NSs.  We work in the body frame $\{\hx_b,\hy_b,\hz_b\}$, where $\hp$ and $\hmf$ are static, with the (here static) rotation axis defining $\hz_b \equiv \hom$, with $\hx_b$ lying in the direction of the projection of $\hp$ onto the plane perpendicular to $\hz_b$.  Because $\hn$ is stationary in the inertial frame, it evolves as $\der \hn/\der t|_b + \bom \btimes \hn = 0$ in the inertial frame, so
\be
\hn(t) = \sin \nu \cos \vphi_n \hx_b - \sin \nu \sin \vphi_n \hy_b + \cos \nu \hz_b,
\label{eq:hn_rot}
\ee
with $\vphi_n(t) = \om t + \vphi_n(0)$ the spin phase.  The spin phase is offset from the rotational phase $\Psi$ (see eq.~[\ref{eq:RVM}]) by $\Psi = \vphi_n + \pi/2 - \Delta_n$, where the offset angle $\Delta_n = \cos^{-1}\left[ (\cos \theta_p - \cos \theta_\om \cos \chi_)/(\sin \theta_\om \sin \theta_p) \right]$.

\subsection{Magnetar undergoing Free Precession}
\label{sec:dyn_freeprec}

The middle panel of Figure~\ref{fig:dyn_models} displays our dynamical model for a freely-precessing NS.  We work in the body frame, where the Cartesian coordinates $\{\hx_b,\hy_b,\hz_b\}$ define the (effective) principal axis of the biaxial NS, with $\hom$ precessing around $\hz_b$ according to
\be
\hom(t) = \sin \theta_\om \cos \vphi_\om \hx_b + \sin \theta_\om \sin \vphi_\om \hy_b + \cos \theta_\om \hz_b,
\label{eq:hom_freeprec}
\ee
where $\vphi_\om(t) = \Omega_{\rm prec} t + \vphi_\om(0)$ is the precession phase, with $\Omega_{\rm prec}$ the precession frequency (see \citealt{ZanazziLai2015,20Zanazzi} for details).  The vector $\hx_b$ lies in the direction of the projection of $\hp$ onto the plane perpendicular to $\hz_b$.

When the precession frequency is much smaller than the rotational frequency ($\Omega_{\rm prec} \ll \om$), $\hom$ remains approximately constant as $\hn$ rotates around $\hom$ over the rotational period of the NS (in the body frame).  Hence $\hn(t)$ is described by an equation similar to~\eqref{eq:hn_rot}, except in a frame where $\hz_b \ne \hom$:
\begin{align}
    \hn(t) = \ &\frac{\sin \nu \cos\vphi_n}{\sin \theta_\om} (\hz_b \btimes \hom) \btimes \hom 
    \nonumber \\
    &- \frac{\sin \nu \sin \vphi_n}{\sin \theta_\om} (\hz_b \btimes \hom) + \cos \nu \hom,
    \label{eq:hn_freeprec}
\end{align}
where $\vphi_n(t) = \om t + \vphi_n(0)$ is the spin phase.

\subsection{Magnetar undergoing Forced Precession}
\label{sec:dyn_forceprec}

The right panel of Figure~\ref{fig:dyn_models} displays our model for forced precession, either by a companion \citep{Yang2020} or a fallback disk \citep{Tong+2020}.  We work in an inertial reference frame $\{\hx_i, \hy_i, \hz_i\}$, with the orbital angular momentum axis of the companion or disk defining $\hz_i$, with $\hx_i$ lying in the direction of the projection of $\hn$ onto the plane perpendicular to $\hz_i$.  Here, the magnetic longitude $\alpha$ is defined as the angle between the planes spanned by the vector pairs $\{\hp,\hom\}$ and $\{\hp,\hmf\}$.
All other quantities are the same as those illustrated in Figure~\ref{fig:setup}.

Assuming the spin angular momentum of the NS ${\bm L} \simeq I \bom$, with $I$ the NS moment of inertia, the NS spin evolves in the forced precession theories according to an equation of the form $\der \hom/\der t|_i - \Omega_{\rm prec} \hz_i \btimes \hom = 0$, hence has a solution
\be
\hom(t) = \sin \theta_\om \cos \vphi_\om \hx_i + \sin \theta_\om \sin \vphi_\om \hy_i + \cos \theta_\om \hz_i,
\label{eq:hom_forceprec}
\ee
where $\vphi_\om(t) = \Omega_{\rm prec} t + \vphi_\om(0)$ is the precession phase.  Because $\hp$ and $\hmf$ are anchored in the rotating NS, they rotate in the inertial frame according to $\der \hp/\der t|_i - \bom \btimes \hp = 0$ and $\der \hmf/\der t|_i - \bom \btimes \hmf = 0$.  Since $\om \gg \Om_{\rm prec}$, $\hom$ stays approximately constant as $\hp$ and $\hmf$ rapidly rotate around $\hom$, hence the motion of $\hp(t)$ and $\hmf(t)$ are approximately described by
\begin{align}
    \hp(t) = & \ \frac{\sin \chi \cos \vphi_n}{\sin \theta_\om} (\hz_i \btimes \hom) \btimes \hom 
    \nonumber \\
    &+ \frac{\sin \chi \sin \vphi_n}{\sin \theta_\om} (\hz_i \btimes \hom) + \cos \chi \hom,
    \label{eq:hp_forceprec} \\
    \hmf(t) = \ &\frac{\sin \chi_f \cos \vphi_n}{\sin \theta_\om} (\hz_i \btimes \hom) \btimes \hom
    \nonumber \\
    &+ \frac{\sin \chi_f \sin \vphi_n}{\sin \theta_\om} (\hz_i \btimes \hom) + \cos \chi_f \hom,
    \label{eq:hmf_forceprec}
\end{align}
where $\vphi_n(t) = \om t + \vphi_n(0)$ is the spin phase.

\section{Application of Periodic Emission Model to FRB 180916}
\label{sec:result}

\begin{figure*}
	\begin{minipage}{0.32\linewidth}
    \includegraphics[width=\linewidth]{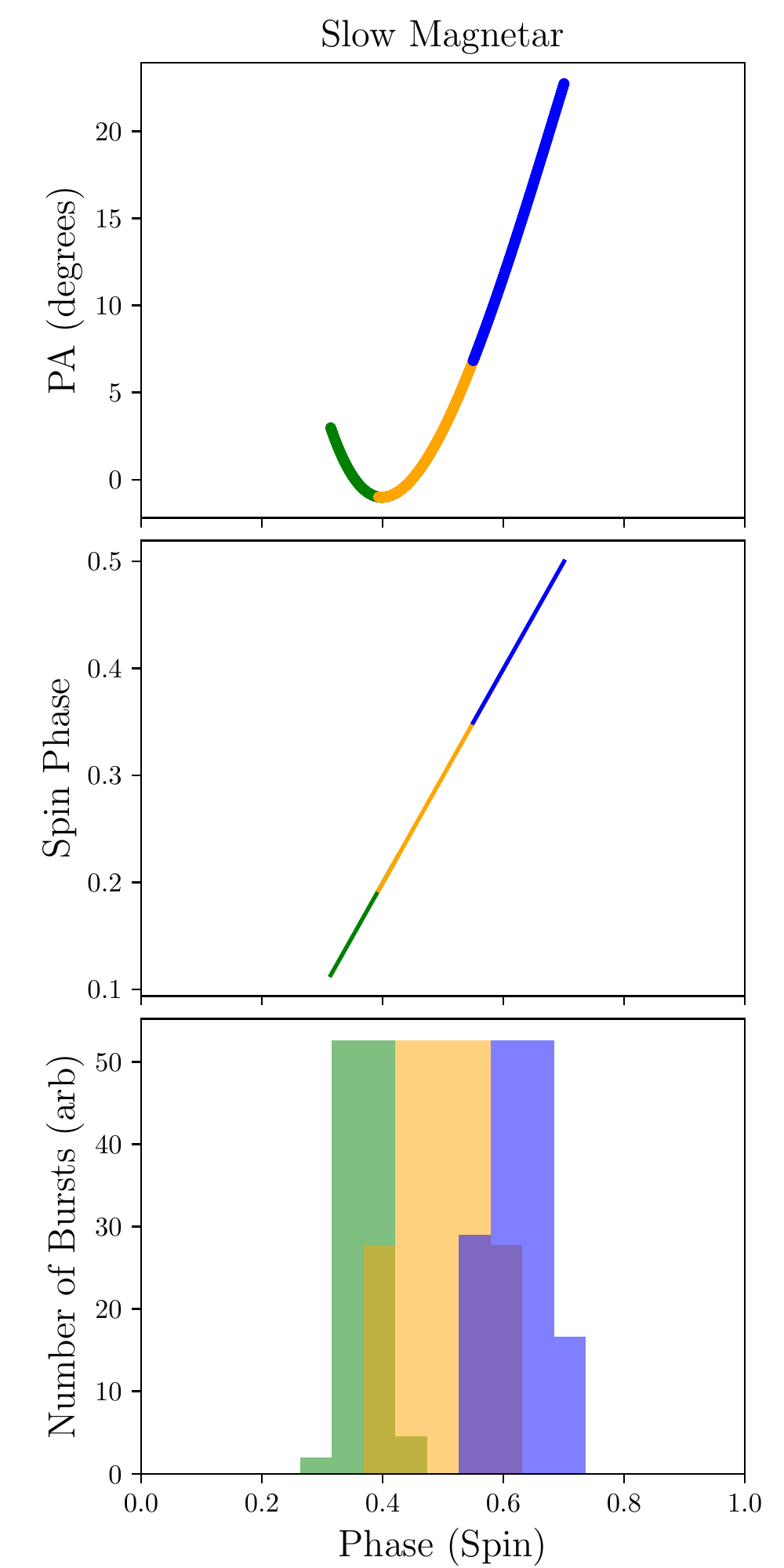}
    \end{minipage}
        \hfill
	\begin{minipage}{0.32\linewidth}
    \includegraphics[width=\linewidth]{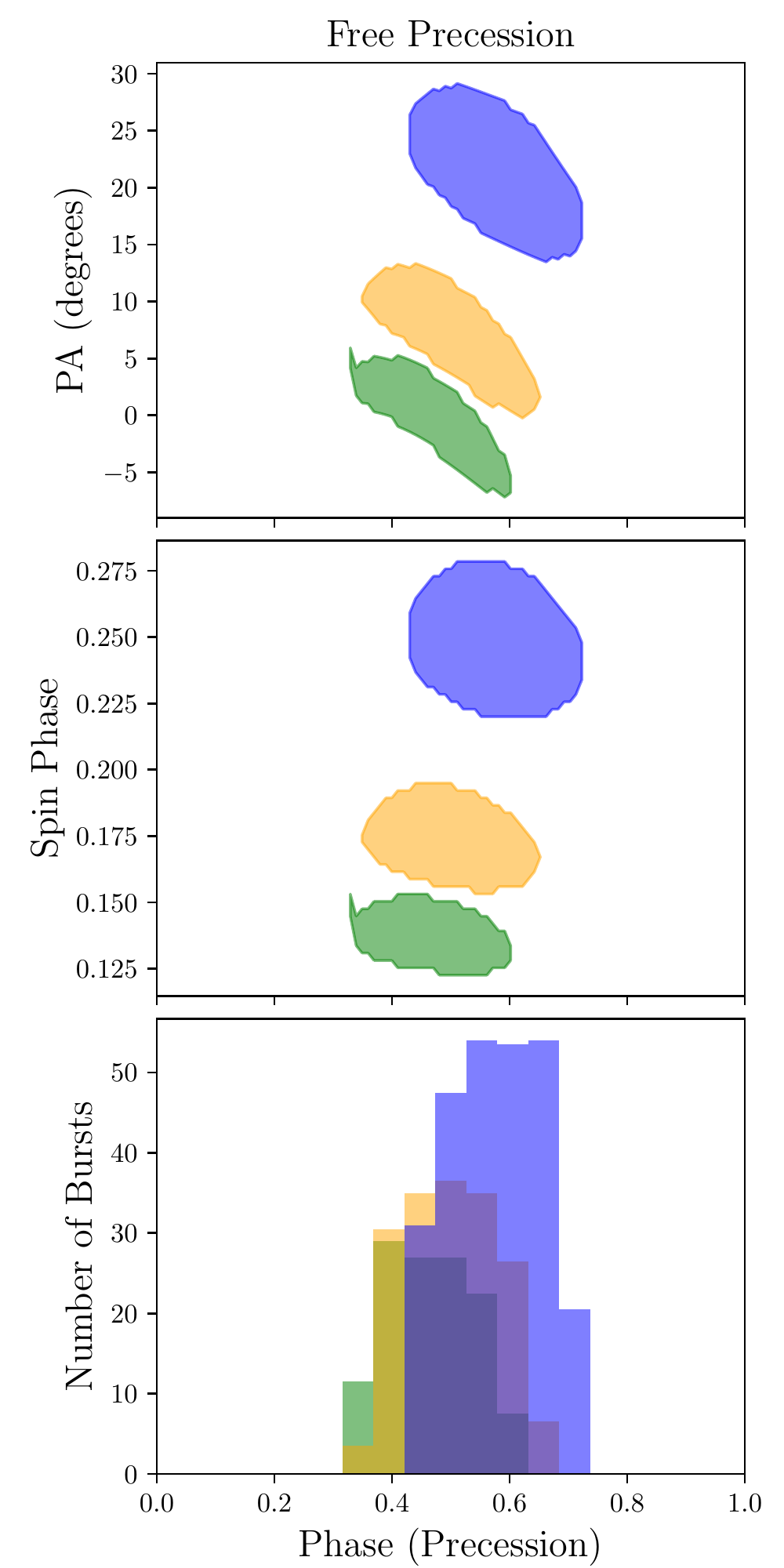}
    \end{minipage}
        \hfill
 	\begin{minipage}{0.32\linewidth}
    \includegraphics[width=\linewidth]{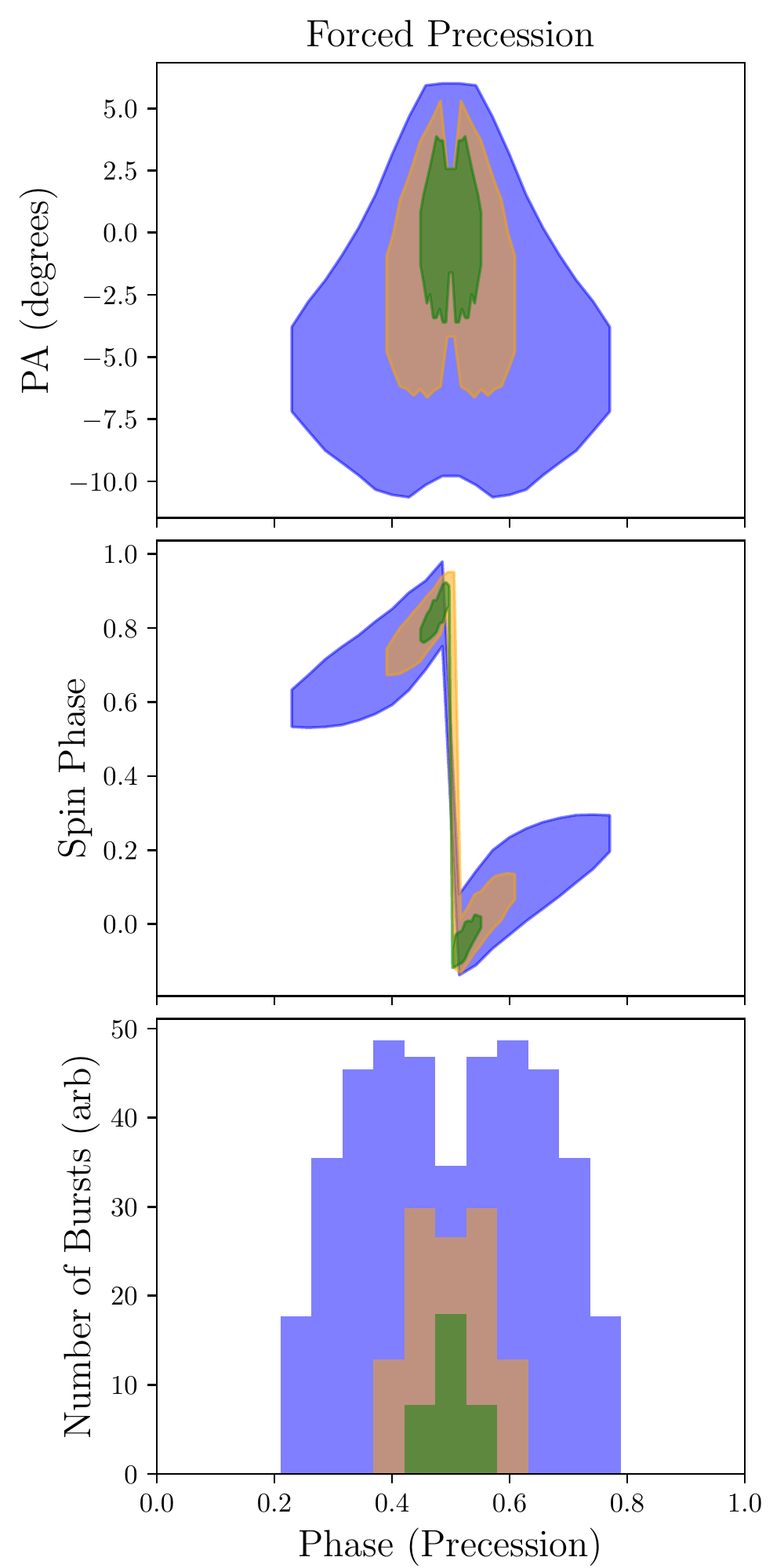}
    \end{minipage}       
        
   \caption{
   Predicted properties of three different models for periodic FRBs: position angle ${\rm PA}$ (top panels), Spin Phase $= \vphi_n/(2\pi)$ (middle panels), and activity windows (bottom panels, Number of Bursts normalization arbitrary), for the frequency bands 1.36-2 GHz (green), 400-800 MHz (orange), and 110-180 MHz (blue).  The observed properties are folded over the timescale which causes the 16.3 day periodicity, which is the rotation period (Phase $=[\vphi_n + \Phi]/[2\pi]$, left panels), or the precession period (Phase $=[\vphi_\om + \Phi]/[2\pi]$, middle \& right panels), with $\Phi$ chosen so that the CHIME band (400-800 MHz) activity window is centered at ${\rm Phase} = 0.5$.  \textit{Left panels:} Emission from a magnetar with slow rotation (\S\ref{sec:dyn_rot}).  Note different frequency bands overlap in PA and Spin Phase.  Fixed model parameters are $\dg_{\rm obs} = 35^\circ$, $45^\circ$, $70^\circ$, and $\xi_f = 5^\circ, 8^\circ$, $20^\circ$, for frequency bands 1.36-2 GHz, 400-800 MHz, 110-180 MHz, respectively, with $\ag = 100^\circ$, $\nu = 40^\circ$, and $\theta_p = 15^\circ$.  \textit{Middle panels:}  Emission from a magnetar undergoing free precession (\S\ref{sec:dyn_freeprec}).  Fixed model parameters are $\dg_{\rm obs} = 35^\circ, 45^\circ, 70^\circ$, and $\xi_f = 5^\circ,8^\circ,20^\circ$, for frequency bands 1.36-2 GHz, 400-800 MHz, 110-180 MHz, repectively, with $\ag = 110^\circ$, $\nu = 40^\circ$, $\theta_\om = 4^\circ$, and $\theta_p = 50^\circ$.  \textit{Right panels:}  Emission from a magnetar undergoing forced precession (\S\ref{sec:dyn_forceprec}).  Fixed model parameters are $\dg_{\rm obs} = 18^\circ,23^\circ,35^\circ$, and $\xi_f = 5^\circ,8^\circ,20^\circ$, for frequency bands 1.36-2 GHz, 400-800 MHz, 110-180 MHz, respectively, with $\ag = 30^\circ$, $\chi = 5^\circ$, $\theta_\om = 25^\circ$, and $\theta_n = 40^\circ$.
   }
   \label{fig:models_obs}
  \end{figure*}
  
Now that we have developed a phenomenological emission model for periodic FRBs due to the motion of NSs, we apply these results to the recent multi-wavelength and polarization measurements of FRB 180916.  For a dynamical model to remain a plausible explanation for the periodicity of FRB 180916, it must explain the following three features of the FRB 180916 emission \citep{20LofarR3,20ApertifR3}:
\begin{enumerate}
    \item The linear polarization angle ${\rm PA}$ varies by less than $\Delta {\rm PA} \lesssim 10^\circ$-$20^\circ$ for single bursts, and $\Delta {\rm PA} \lesssim 40^\circ$ between bursts.
    \item The bust activity window (region in phase busts seen after folding over 16.3 day periodicity) widens with a decrease in frequency.
    \item The burst activity windows in different frequency bands have phase centers which are offset from one another, or busts at one frequency are systematically delayed with respect to burst at another frequency.  We will call this effect activity window phase drift with frequency.  
\end{enumerate}

The first constraint (with $\Delta {\rm PA}$) lied at odds with the original predictions of the \cite{20Zanazzi} free precession model.  This is because the emission region was assumed to emit in the direction of the NS dipole moment ($\hmf = \hp$, see Fig.~\ref{fig:setup}).  Because FRBs occur when $\hn \approx \hmf = \hp$, the rotational phase $\Psi \ll 1$ during an FRB, implying equation~\eqref{eq:RVM} reduces to
\be
\tan {\rm PA} \simeq \frac{\sin \chi}{\sin (\chi - \nu)} \Psi.
\label{eq:PA_red}
\ee
Since $\hn \approx \hp$ implies $\chi \approx \nu$ (see Fig.~\ref{fig:setup}), the denominator of equation~\eqref{eq:PA_red} becomes large, and hence the model of \cite{20Zanazzi} predicted large variations in ${\rm PA}$ during individual FRB bursts, as well as modulation of the ${\rm PA}$ variation between bursts as the NS precessed.  However, because $\hmf$ and $\hp$ can have significant differences in orientation (see \S\ref{sec:emit_dir}), it is possible for ${\rm PA}$ variations to be sufficiently small to be consistent with the polarization measurements from FRB 180916.

In this section, we calculate the variations in PA, as well as how the burst activity window depends on frequency $f$, with different dynamical models (Fig.~\ref{fig:dyn_models}).  We create maps of the range in PA variation, and estimate the activity window of different models, by first fixing parameters which are constant for a given dynamical model (see Fig.~\ref{fig:setup}), and in particular specifying $\hmf$ and $\xi_f$ for different frequency bands. Notice here, we assume the shift of emission region in the observed polar angle $\delta_\mr{ob}$ while the center of the emission region has the same magnetic longitude $\alpha$.  A burst is considered observed at a specific $f$-band if $\hn$ and $\hmf$ lie within an angle $\xi_f$ of each other, with no emission if this condition is not met.  The Phase $= (\vphi_n + \Phi)/(2\pi)$ for the slow magnetar model, Phase $= (\vphi_\om + \Phi)/(2\pi)$ for the free/forced precession model (with $\Phi$ a constant picked so the 400-800 MHz $f$-band is centered at Phase $=0.5$), and Spin Phase $= \vphi_n/(2\pi)$ are then cycled through their possible parameter values ($\vphi_\om, \vphi_n \in [0,2\pi]$, middle panels of Fig.~\ref{fig:models_obs}) to calculate the range of PA values with (spin or precession) Phase (top panels of Fig.~\ref{fig:models_obs}).  The number of bursts over all spin phases are then binned by precessional phase to construct burst activity window profiles (bottom panels of Fig.~\ref{fig:models_obs}).

The top panels of Figure~\ref{fig:models_obs} display the PA variation with either spin (left panels) or precession (middle \& right panels) phase.  From this, we see all three models are able to have PA variations consistent with observations, as long as the emission region direction $\hmf$ has a significant offset from the dipole axis $\hp$ ($\dg_{\rm obs}$ sufficiently large).  The spin phase has similar variations as the PA with spin or precession phase, with differences due to the slight differences in geometry between the two angles (see eq.~[\ref{eq:RVM}]).

The bottom panels of Figure~\ref{fig:models_obs} display the activity window over different $f$-bands.  The magnetar with a slow rotation period, as well as the magnetar undergoing free precession, can accommodate an activity window which widens at lower $f$ values, as well as activity window phase drifts between $f$-bands.  The widening of the activity window is primarily due to the increase of $\xi_f$ at lower $f$-bands, while the phase drift is primarly due to the higher magnetic polar angle $\dg_{\rm obs}$ values at lower $f$-bands (see \S\ref{sec:emit_dir} for discussion).  The roughly monochromatic bust rate with spin phase for the slow magnetar model is because the event rate is assumed to be uniform over the NS spin phase: dropping this assumption can lead to activity windows which are not monochromatic.  We conclude that the dynamical model of a magnetar with a slow rotation period, as well as a magnetar undergoing free precession, are capable of causing periodic emission consistent with observations of FRB 180916.

The bottom right panel of Figure~\ref{fig:models_obs} displays the activity windows for the model describing a magnetar undergoing forced precession.  Although these parameters clearly show a widening activity window with a lower $f$-band, the histogram also displays no activity window phase shift with frequency.  The lack of activity window phase shift is not unique to these particular model parameters, but is rather a general feature of the forced precession model.  Consider a magnetar undergoing forced precession, with LOS direction $\hn = \sin \theta_n \hx_i + \cos \theta_n \hz_i$.  Because $\hn \bcdot \hmf (\vphi_\om,\vphi_n) = \hn \bcdot \hmf(-\vphi_\om,-\vphi_n)$ (see eqs.~[\ref{eq:hom_forceprec}] \&~[\ref{eq:hmf_forceprec}]), the activity window is always symmetric about $\vphi_\om = \vphi_n = 0$, irrespective of the emission frequency $f$.  
Hence, the observed activity window phase drift with $f$ in FRB 180916 disfavors the dynamical model of a magnetar undergoing forced precession.

\section{Proposed Observational Tests}
\label{sec:obs}

The previous section showed the dynamical model of a magnetar with a slow rotation period, and a magnetar undergoing free precession, could explain the polarization and frequency-dependent activity window of FRB 180916, while the magnetar undergoing forced precession was disfavored.  In this section, we describe how further polarization and activity window measurements can refine constraints on model parameters, and additional observations can favor or rule-out the remaining dynamical theories to explain periodic FRBs.

If additional periodic FRB sources are detected, the PA swings and activity windows are expected to change from source to source.  Within the context of our phenomenological emission model, the phase-drift of the FRB activity window with frequency results from a change of emission region.  This phase-drift can be in either direction, depending on the direction of rotation/precession, as well as the orientation of the emission regions with respect to each-other.  Event rates of FRBs can also differ significantly across frequency due to different magnetic field geometries. However, both the slowly-rotating and freely-precessing magnetars require the emission regions to be asymmetric with respect to the NS dipole axis.  In contrast, it is difficult for the induced precession model to have an activity window which drifts asymmetrically in phase with frequency.

All dynamical models allow a small PA change with precession or spin phase, while larger PA changes like in the case of FRB 180301\citep{Luo+2020} is also allowed. The PA variations of both models differ between frequency bands. 
However, for a slowly-rotating magnetar, there is a constant PA angle at each (spin) phase, while for the freely-precessing magnetar,  a range of PA values are expected at each (precession) phase (Fig.~\ref{fig:models_obs}, upper panels).  Moreover, for a fixed precession phase, only certain spin phases will lead to an observed burst (Fig.~\ref{fig:models_obs}, middle panels).  More accurate PA measurements from high S/N bursts would be able to distinguish between these two models (or rule out both). 
 
The precession model also requires fixed emission region, and a small timescale periodicity due to the rotation period of the magnetar. This short periodicity may be concealed by various timing noises common in young magnetars. However, if a short period is found, the precession model predicts a change of the short periodicity duty cycle with the NS precession phase. For instance, in Figure~\ref{fig:models_obs} central middle panel, the spin duty cycle appears smaller at the edge of each active (precession) phase.  Moreover, the active window in the spin phase is changing against frequency with this setup of asymmetric emission against magnetic pole. 

To get the same activity windows, the free-precession model requires a smaller emission region than the slowly-rotating magnetar model. This is because for each spin phase of the slowly-rotating magnetar, the LOS points at a specific location on the NS surface, while for each precession phase of the freely-precessing magnetar, the LOS rotates around the NS rotation axis.  Distinguishing between these two models would constrain the FRB emission region altitude and angular size.

Recently, \cite{Katz2020} discussed how detecting a change in the FRB period can further differentiate between dynamical theories.  We note the freely-precessing magnetar model expects a much larger period change than the slowly-rotating magnetar model, since the spin-down timescale is much shorter in the former model due to the faster rotation frequency \cite[e.g.][]{20Zanazzi}.

\section{Summary and Conclusions}
\label{sec:conc}
In this paper, we construct a phenomenological model of Fast Radio Burst (FRB) emission from rotating magnetars to test periodic FRB theories, in light of recent measurements of small polarization Position Angle (PA) swings during and between bursts, as well as the narrowing and phase shift of burst activity windows with frequency.  Our model assumes burst are emitted from a region anchored into the magnetar and offset from the dipole axis, with region size and dipole offset angle increasing with a decrease in frequency, due to the altitude dependence of curvature radiation on frequency (\S\ref{sec:emit_dir}).  The model PA values are given by the rotating vector model (\S\ref{sec:emit_pol}).  

Using this model, we constrain three separate dynamical models which have been invoked to explain the 16.3 day periodicity of FRB 180916: a magnetar with a slow rotation period (\S\ref{sec:dyn_rot}), a magnetar undergoing free precession (\S\ref{sec:dyn_freeprec}), and a magnetar undergoing forced precession due to an external body (\S\ref{sec:dyn_forceprec}).  We find the slowly-rotating and freely-precessing magnetar models can produce PA swings and frequency-dependent activity windows consistent with recent observations, but the magnetar undergoing forced precession is disfavored as a dynamical model, due to its inability to produce a frequency-dependent phase drift of the burst activity window (\S\ref{sec:result}).  Future observations are necessary to distinguish between the remaining theories, and understand what is causing the periodicity of FRB 180916 (\S\ref{sec:conc})


\section*{Acknowledgements}

DL acknowledges the discussion with Liam Connor and Vikram Ravi.  JZ is supported by a CITA postdoctoral fellowship.


\bibliographystyle{aasjournal}

\end{document}